\documentclass[twocolumn,prb,showpacs,multicol,amsmath,amssymb]{revtex4-1}
\usepackage{graphicx}
\usepackage{epstopdf}
\newcommand{\be}{\begin{equation}}
\newcommand{\ee}{\end{equation}}

\newcommand{\bea}{\begin{eqnarray}}
\newcommand{\eea}{\end{eqnarray}}
\newcommand{\bd}{\begin{displaymath}}
\newcommand{\ed}{\end{displaymath}}
\newcommand{\ba}{\begin{array}}
\newcommand{\ea}{\end{array}}
\newcommand{\bi}{\begin{itemize}}
\newcommand{\ei}{\end{itemize}}
\newcommand{\bc}{\begin{center}}
\newcommand{\ec}{\end{center}}
\newcommand{\bfl}{\begin{flushleft}}
\newcommand{\efl}{\end{flushleft}}
\newcommand{\bfr}{\begin{flushright}}
\newcommand{\efr}{\end{flushright}}

\def\bk{{\bf k}} \def\bq{{\bf q}}

\def\6{\partial}

\def\={\!\!\!&=&\!\!\!}
\def\+{\!\!\!&&\!\!\!+~}
\def\-{\!\!\!&&\!\!\!-~}

\begin{document}
\date{\today}
\title{Quasiparticle interference in heavy Fermion superconductor $\rm{ CeCoIn_5}$}

\author{Alireza Akbari$^{1}$}
\author {Peter Thalmeier$^{1}$}
\author {Ilya Eremin$^{2}$}

 \affiliation{
   $^1$Max Planck Institute for the  Chemical Physics of Solids, D-01187 Dresden, Germany
\\ $^{2}$ Institut f\"ur Theoretische Physik III, Ruhr-Universit\"{a}t Bochum, D-44780, Bochum, Germany
 }

 \begin{abstract}
We investigate the quasiparticle interference in the heavy Fermion superconductor $\rm{ CeCoIn_5}$ as direct method
to confirm the d-wave gap symmetry. The ambiguity between $d_{xy}$ and $d_{x^2-y^2}$ symmetry remaining from earlier specific heat and thermal transport investigations has been resolved in favor of the latter by the observation of a spin resonance that can occur only in $d_{x^2-y^2}$ symmetry. However these methods are all indirect and depend considerably on theoretical interpretation. Here we propose that quasiparticle interference (QPI) spectroscopy by scanning tunneling microscopy (STM) can give a direct fingerprint of the superconducting gap in real space which may lead to a definite conclusion on its symmetry for $\rm{ CeCoIn_5}$ and related 115 compounds. The QPI pattern for both magnetic and nonmagnetic impurities is calculated for the possible d-wave symmetries and characteristic differences are found that may be identified by STM method.
 \end{abstract}


\maketitle

%%%%%%%%%%%%%%%%%%%%%%%%%%%%%%%%%%%%%%%%%%%%%%%%%%%%%%%%%%%%%%%%%%%%%%%%%%
%%%%%%%%%%%%%%%%%%%%%%%%%%     Section I      %%%%%%%%%%%%%%%%%%%%%%%%%%%%
%%%%%%%%%%%%%%%%%%%%%%%%%%%%%%%%%%%%%%%%%%%%%%%%%%%%%%%%%%%%%%%%%%%%%%%%%%
\section{Introduction}

In recent years the spectroscopic imaging  scanning tunneling microscopy (SI-STM) has become a powerful
experimental tool for studying the local electronic properties of various
superconductors \cite{Yazdani97,Hoffman02, Balatsky06}.
It is well known that the Fourier  transform of STM (FT-STM) data or
quasiparticle interference (QPI),
can be used to  elucidate the nature of the many-body states
in novel superconductors,
in particular those having quasi-two-dimensional electronic structure.
In the presence of impurities, elastic scattering mixes two quasiparticle eigenstates
with momenta {\bf k}$_1$ and {\bf k}$_2$ on a
contour of constant energy. The resulting interference at wave vector ${\bf q} = {\bf k}_2 - {\bf k}_1$
reveals a modulation of the local density of states (LDOS). The
interference pattern in momentum space can
be visualized by means of the SI-STM\cite{mkelroy,Sprunger97}.

In layered cuprates the analysis of the QPI
has provided details of the band structure,
the nature of the superconducting gap, or other competing
orders\cite{cuprates,cuprates1,cuprates2,cuprates3}.
It was shown that a magnetic-field dependence in quasiparticle scattering interference patterns is
sensitive to the sign of the anisotropic gap\cite{hanaguri_cup, coleman}.
Recently, such QPI effects have also been studied in the iron based superconductors\cite{Kamihara08},
to find the order-parameter symmetry\cite{Chuang10, hanaguri10, Zhang09,akbari} in a case
where we have multi-band superconductivity
with possible sign reversal of the order parameter  between the electron and hole pockets.

%%%%%%%%%%%%%%%%%%%%%% figure %%%%%%%%%%%%%%%%%%%%
\begin{figure}[t]
\centerline{
\includegraphics[width=0.55\linewidth]{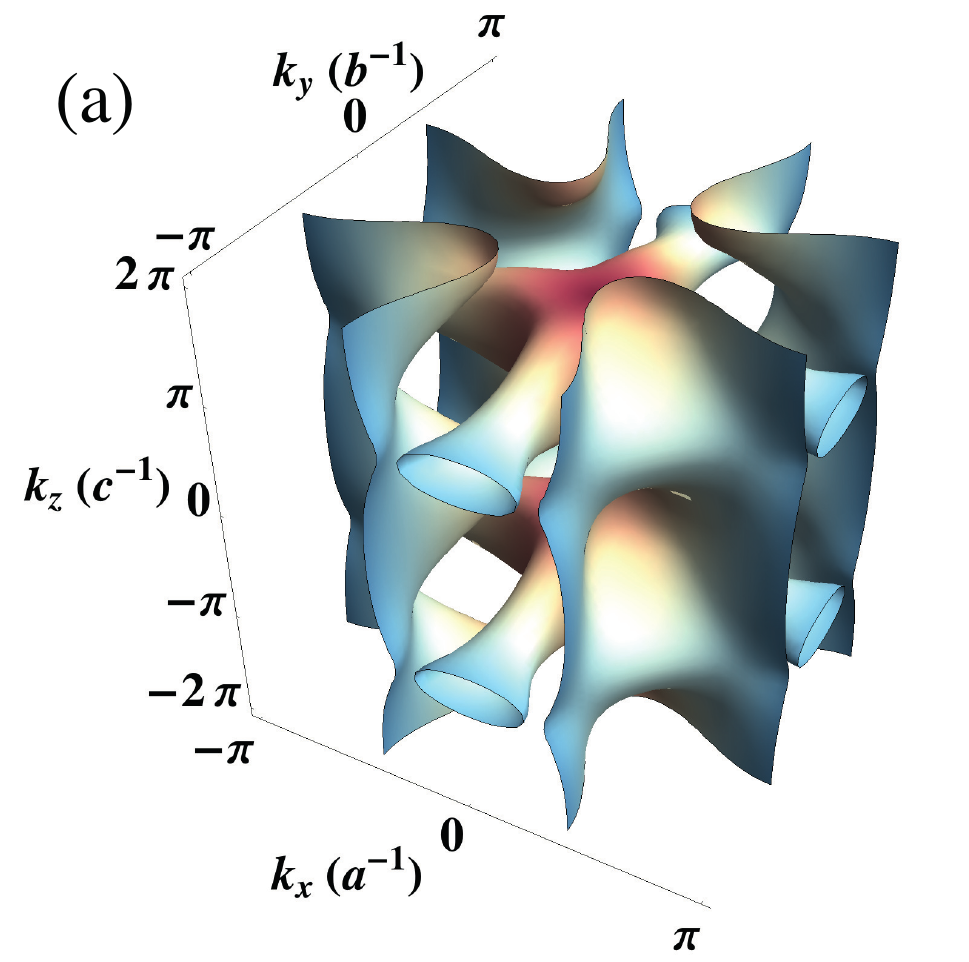}
}
\centerline{
\includegraphics[width=0.95\linewidth]{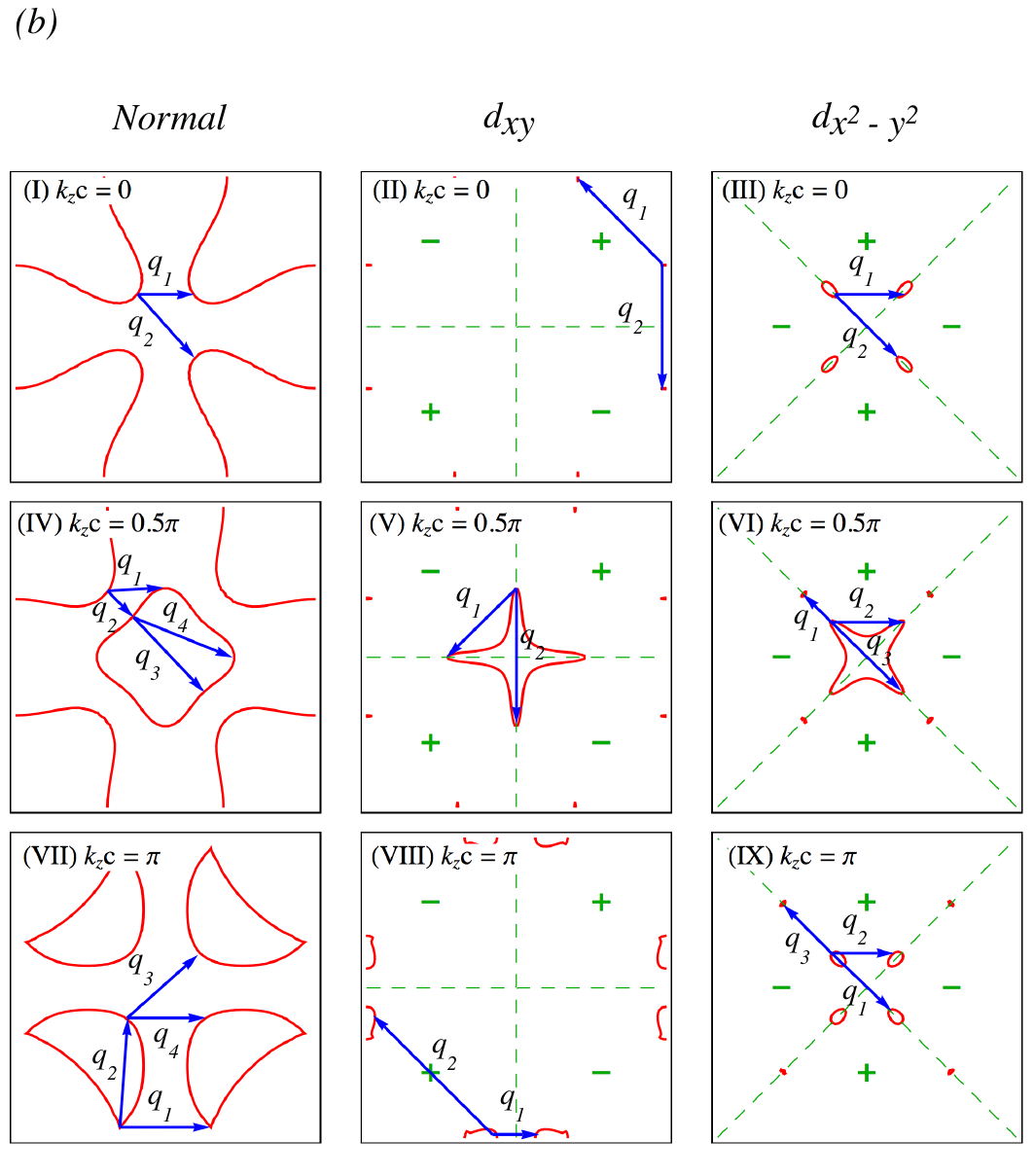}
}
\caption{(Color online)  a) Calculated Fermi surface for $\rm{ CeCoIn_5}$ using the band structure parameters defined in Ref.~[\onlinecite{tanaka:06}],
b) first panel indicates a cut through the Fermi surface for normal, and second and third panel indicate a cut through the Fermi surface for  $d_{xy} $ and $d_{x^2-y^2}$ gap symmetry respectively (solid red lines)  for bias voltage  $\omega=0.1\Delta_0$.
(at $k_z=0$ first row, at $k_z=0.5\pi$ second row, and at $k_z=\pi$ last row).
Dashed (green) lines indicate the node-lines,
$\pm$ denotes the sign of the superconducting  gap,
and $\bq_i$ are the typical scattering vectors defining the QPI pattern.}
%}
\label{Fermi}
\end{figure}
%%%%%%%%%%%%%%%%%%%%%%fig%%%%%%%%%%%%%%%%%%%%%%%%%

Applying  this technique  in the heavy Fermion systems and considering QPI in these materials is one of great interest, in particular because there are numerous unconventional superconductors
with competition of magnetism and superconductivity as well as non-Fermi liquid behaviour.
One of the most difficult issues in the heavy Fermion superconductors like the 115 compounds  $\rm{ CeMIn_5}$  ($\rm{ M = Co, Ir, Rh}$)  is the identification of the symmetry of superconducting order parameter.
In the present context of QPI theory the microscopic origin of the unconventional superconductivity is not an issue. For $\rm{ CeCoIn_5}$  various proposals based on spin fluctuation theory \cite{takimoto:02,kubo:06} and composite pairing \cite{flint:11} have been advanced.

Commonly several candidates for the gap function are proposed and their compatibility with temperature and field dependence of thermodynamic and transport quantities is used to discriminate between them \cite{matsuda:06, thalmeier:05}. In this respect it is instructive to recall the previous discussions on the gap symmetry  in this compound.
Firstly the Pauli limiting behaviour of the upper critical field \cite{tayama:02} and observed Knight \cite{kohori:01} shift proves the spin singlet nature of the gap suggesting d-wave pairing.
Originally field-angle resolved  thermal conductivity \cite{izawa:01} and specific heat \cite{aoki:04} experiments which probe the node structure gave conflicting results of $d_{x^2-y^2}$ and  $d_{xy}$  gap symmetries respectively. Then the observation of a pronounced spin resonance \cite{stock:08}
at $\omega_r/2\Delta_0=0.65$ in the superconducting state with inelastic neutron scattering (INS) gave strong evidence for the former. Namely detailed calculations of the spin response with realistic Fermi surface \cite{eremin:08} show that the resonance can appear only for the $d_{x^2-y^2}$ symmetry but not in the $d_{xy}$ case. Further field-angle resolved specific heat measurements at even lower temperature \cite{an:10} finally also concluded on $d_{x^2-y^2}$ gap symmetry. These interpretations however all depend considerably on theoretical model features and approximations.
For example, the quasiparticle relaxation rate in the vortex state for the transport properties influences the results of field-angle resolved specific heat measurements. In addition, the results of the INS experiments were also interpreted in terms of spin wave excitations which sharpen in the superconducting state due to effect of the gap on the normal state Landau damping.\cite{chubukov:08}
It would be preferable to have a more direct method that can provide a fingerprint of superconducting gap symmetry. In this work we show that QPI can indeed serve this purpose
in heavy Fermion superconductor  $\rm{ CeCoIn_5}$ and other related 115 systems where the degree of the three-dimensionality is more substantial than in cuprates or iron-based superconductors.
In Sec. \ref{model} we present the theoretical framework of this model. Then Sec. \ref{numerical} presents the numerical results for the QPI and how to interpret them in terms of Fermi surface and nodal gap properties of the proposed order parameter candidates.
Finally Sec.~\ref{summary} gives a brief summary and outlook.

%%%%%%%%%%%%%%%%%%%%%%%%%%%%%%%%%%%%%%%%%%%%%%%%%%%%%%%%%%%%%%%%%%%%%%%%%%
%%%%%%%%%%%%%%%%%%%%%%%%%%     Section I      %%%%%%%%%%%%%%%%%%%%%%%%%%%%
%%%%%%%%%%%%%%%%%%%%%%%%%%%%%%%%%%%%%%%%%%%%%%%%%%%%%%%%%%%%%%%%%%%%%%%%%%
\section{Theoretical Model}
\label{model}

As a starting point we need a model for the electronic structure which captures the essence of the heavy quasiparticle bands and their Fermi surface but is technically still manageable for a T-matrix calculation of QPI in the superconducting state. The electronic structure of $\rm{ CeMIn_5}$ has been investigated by using tight binding models with hybridisation for
f-electrons and p-(conduction) electrons \cite{maehira:03,tanaka:06}. The Ce-4f electron states are split by a large spin orbit-coupling ($\Delta_{so}\sim$ 0.4 meV \cite{maehira:03}) into upper j=7/2 and lower j=5/2 multiplets. One may therefore restrict to the lower one as done in Ref.~\onlinecite{maehira:03} which is further split into three crystalline electric field (CEF) Kramers doublet states. Because the CEF splitting energy is about three times the heavy quasiparticle band width ($W\simeq$ 4 meV) one may further restrict to the lowest CEF doublet \cite{tanaka:06} which has an effective pseudo-spin 1/2. Then the Anderson lattice model Hamiltonian  for the two
hybridized (c,f) conduction and localized orbitals which are doubly spin degenerate is given by
\begin{eqnarray}
{\cal H}
&=&
\sum\limits_{{\bf k}\sigma }
\epsilon^c_{{\bf k}}c_{{\bf k}\sigma
}^{\dagger}c_{{\bf k}\sigma }
+
\epsilon^f_{{\bf k}} f_{{\bf k}\sigma} ^{\dagger}f_{{\bf k}\sigma}
+V_{{\bf k}}\left( c_{{\bf k}\sigma }^{\dagger}f_{{\bf k}\sigma}
+h.c.\right)
 \nonumber \\
&&+\sum\limits_{{\bf k} {\bf k}^\prime } U_{ff}f_{{\bf k}\uparrow }^{\dagger}f_{{\bf k}\uparrow }f_{{\bf k}^\prime\downarrow}^{\dagger}f_{{\bf k}^\prime\downarrow }.
\label{eq-}
\end{eqnarray}
where $c_{{\bf k}\sigma
}^{\dagger}$ creates an electron with spin $\sigma$
in the conduction orbital  with wave vector ${\bf k}=(k_x,k_y,k_z)$.
Furthermore, $\varepsilon^c_{{\bf k}}$ and $\varepsilon^f_{{\bf k}}$  are effective tight binding
dispersions of the conduction band and the renormalized dispersion for the $f$ band respectively.
$f_{{\bf k}\sigma} ^{\dagger}$ creates the f electron with momentum \bk~ and pseudo spin $\sigma$, and $U_{ff}$ is its on-site Coulomb repulsion. Finally $V_{{\bf  k}}$ is the  hybridization energy between the lowest 4f doublet
and conduction bands which contains implicitly the effect of spin orbit and CEF term. A use the Anderson lattice model of Ref.~\onlinecite{tanaka:06} is preferable here because it provides a convenient way to realistically model the heavy quasiparticle bands. For complementarity we mention that  there is also an alternative method to investigate SI-STM, proposed in Refs.~\onlinecite{maltseva:09,yuan:11}, which starts  from the Kondo lattice model where charge fluctuations are eliminated and 4f electrons are considered as fully localized.

%%%%%%%%%%%%%%%%%%%%%% figure %%%%%%%%%%%%%%%%%%%%
\begin{figure}[!Ht]
\centerline{
\includegraphics[width=1\linewidth]{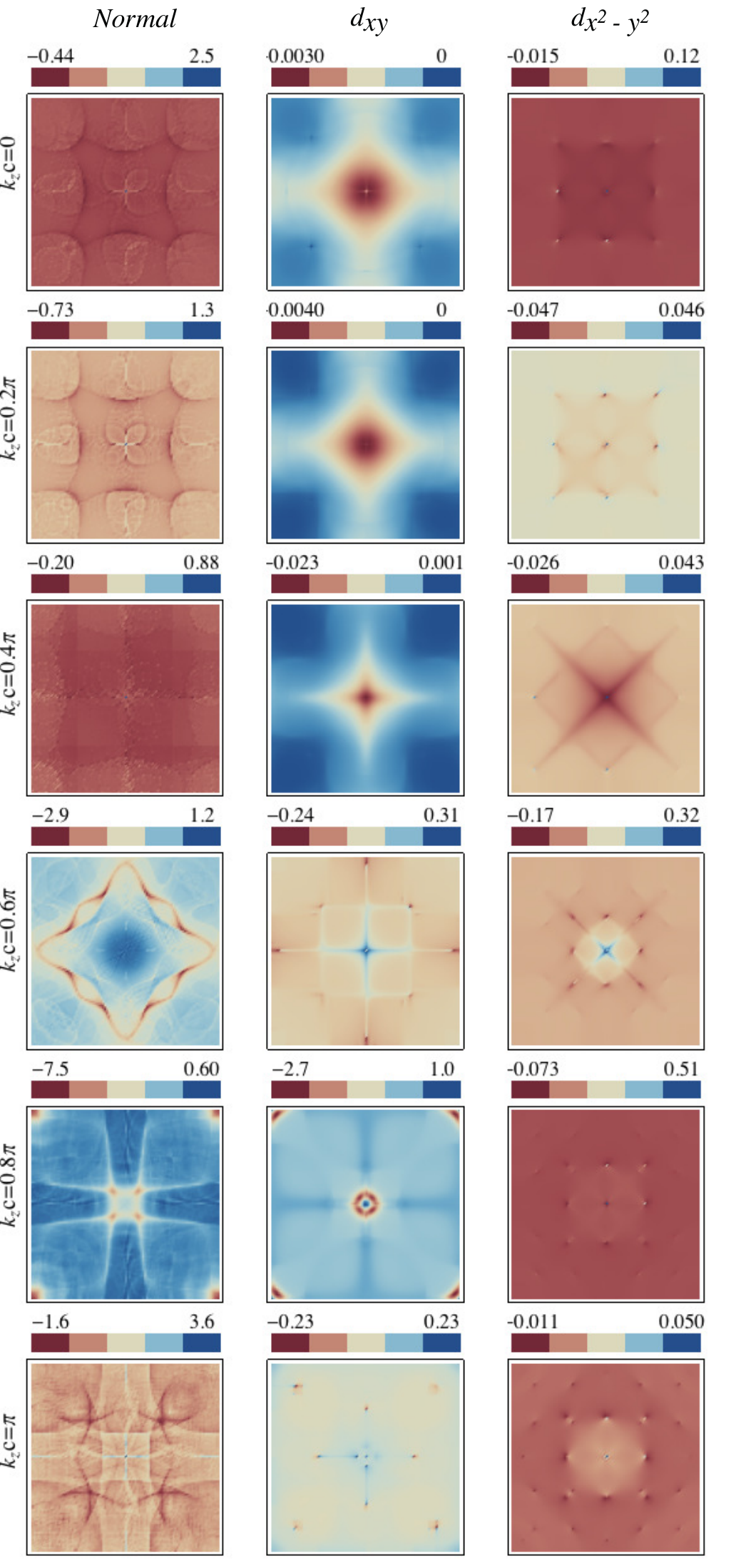}
}
\caption{(Color online)
The Quasiparticle interference  due to non-magnetic impurity impurity scattering,
for individual layers at
 $k_zc=0.0,0.2\pi,0.4 \pi, 0.6 \pi, 0.8 \pi, \pi$,
 for normal state (first panel),
 superconducting state with $d_{xy}$ gap  symmetry (second panel) and  $d_{x^2-y^2}$ gap symmetry (third panel).
 (For small bias voltage  $\omega=0.01\Delta_0$ and using $\Delta_0/W=0.086$ where W is the total quasiparticle band width.).}
%}
\label{Difkz}
\end{figure}
%%%%%%%%%%%%%%%%%%%%%%fig%%%%%%%%%%%%%%%%%%%%%%%%%

It is known that\cite{tanaka:06} in  the limit $U_{ff}\rightarrow \infty$ where double occupation  of the f-states are excluded,
by defining the auxiliary boson one can find the mean field  the Hamiltonian as
\bea
{\cal H}_{MF}&=& \sum\limits_{{\pm},{\bf k}\sigma }E^{\pm}_{{\bf k}}a^{\dagger}_{\pm,{\bf k}\sigma}a_{\pm,{\bf k}\sigma},\nonumber\\
E^{\pm} _{\bf  k}&=&\frac{1}{2}
\biggl[
\epsilon^{c}_{{\bf  k}}+\epsilon^{f}_{{\bf  k}}\pm\sqrt{(\epsilon^{c}_{{\bf  k}}-\epsilon^{f}_{{\bf  k}})^2+4\tilde{V}^2_{{\bf  k}}}
\biggr].
\eea
where $E^{,\pm} _{\bf  k}$ are the hybridized quasiparticle ($a_{\pm,{\bf k}\sigma}$) bands with
$\tilde{V}^2 _{\bf  k}=V^2 _{\bf  k}(1-n_f)$ denoting the effective hybridisation obtained by
projecting out double occupancies. Due to $1-n_f\ll 1$ $\tilde{V} _{\bf  k}$ is strongly reduced
with respect to the single particle $V _{\bf  k}$.
Using the parameters defined in Ref.~[\onlinecite{tanaka:06}] for the above quasiparticle band structure
we plot the corresponding Fermi surface (FS) in Fig.~(\ref{Fermi}.~a).
The FS crossing originates only from one of the bands, $E^{-} _{\bf  k}$, while $E^{+} _{\bf  k}$ remains well above the Fermi energy. Therefore, in the following analysis of the QPI it does not play a significant role.

To supplement the normal state Hamiltonian we add the superconducting  pairing term given by
\bea
{\cal H_{SC}}=\sum\limits_{{\bf k}}\Delta_{{\bf k}}\left(
a_{\pm,{\bf k}\uparrow }^{\dagger}a_{\pm,-{\bf k}\downarrow }^{\dagger}+h.c.\right)
\eea
where $\Delta_{{\bf k}}$ is the superconducting gap function.

The scattering of the quasiparticle by
perturbations in the sample such as non-magnetic or magnetic impurities is
responsible for the QPI which is believed to be measured in SI-STM . We perform the
analysis of such processes based on a T-matrix description\cite{Balatsky06}. In particular,  we introduce an
impurity term in the Hamiltonian
\bea
{\cal H}_{imp}=\sum\limits_{{\bf k} {\bf  k}^\prime\sigma\sigma^\prime}
\left(
J^0_{ {\bf  k}{\bf  k}^\prime}
\delta_{\sigma\sigma^\prime}+ J_{\sigma\sigma^\prime }{\bf S}\cdot {\bf \sigma}_{\sigma\sigma^\prime}
\right)
a^{\dagger}_{\pm,{\bf k}\sigma}a_{\pm,{\bf k}^\prime\sigma^\prime},
\nonumber\\
\eea
where $J^0_{ {\bf  k}{\bf  k}^\prime}$ and  $J_{\sigma\sigma^\prime }$
represent the non-magnetic and the magnetic  point-like
scattering between the electrons respectively. Note that in this picture Ce does no longer
carry a local magnetic moment because below the coherence temperature its f-electron is incorporated in the
itinerant quasiparticle states. The magnetic impurities may be other 4f ions with stable moment. Depending on
the crystalline electric field splitting their coupling to quasiparticle states can become strongly anisotropic,
e.g. of Ising type in the simplest case which will be considered in the following. This means we set the quantization axis of the magnetic impurity spin along the $z$-direction and consider only the $S_z$ spin component. Here, {\bf S} refers again to the pseudo spin of the lowest Kramers doublet of the 4f impurity moment.  At this stage we do not include scattering of quasiparticles by nearly critical collective spin fluctuations which may be important in $\rm{ CeCoIn_5}$ because of its closeness to an antiferromagnetic quantum critical point \cite{paglione:03}.

Defining  the new Nambu spinor as
$
\hat{\psi}_{{\bf k}}^{\dagger}=(
a^{\dagger}_{+,{\bf k} \uparrow},
a_{+,-{\bf k} \downarrow},
a^{\dagger}_{-,{\bf k} \uparrow},
a_{-,-{\bf k} \downarrow}
)
$,
the Hamiltonian can be written as
\begin{eqnarray}
{\cal H} =
\sum\limits_{{\bf k} }
\hat{\psi}_{{\bf k} }^{\dagger}\hat{\beta}_{{\bf k}}
\hat{\psi}_{{\bf k} }
+
\sum\limits_{{\bf k}{\bf  k}^\prime }
\hat{\psi}_{{\bf k} }^{\dagger}\hat{U}_{ {\bf  k}{\bf  k}^\prime}
\hat{\psi}_{{\bf  k}^\prime}
\end{eqnarray}
By introducing $J^0_{ {\bf  k}{\bf  k}^\prime}=\gamma $;
and $J_{zz}S_z=\gamma^\prime $, the above matrices
$\hat{\beta}_{{\bf k}}$ and $\hat{U}_{ {\bf  k}{\bf  k}^\prime}$ are  defined as
\begin{eqnarray}\nonumber
\hat{\beta}_{{\bf k}}= &&
(\frac{\tau_0+\tau_z}{2})
\otimes
(E^{+} _{\bf  k}\sigma_z+\Delta_{{\bf k}} \sigma_x)
\\
&+&(\frac{\tau_0-\tau_z}{2})
\otimes
(E^{-} _{\bf  k}\sigma_z+\Delta_{{\bf k}} \sigma_x);
\eea
and
\bea
 \nonumber \\&& \hspace{0.25cm}
\hat{U}_{ {\bf  k}{\bf  k}^\prime}=
\tau_0
\otimes
(\gamma^\prime\sigma_0+\gamma \sigma_z).
\end{eqnarray}
Here $\sigma_i$ are the Pauli matrices acting in spin space,  $\tau_i$  are the Pauli matrices in the orbital space, and $\tau_i\otimes\sigma_i$  denotes a
direct product of the matrices operating on the 4-dimensional Nambu space.

Therefore in terms of the Nambu spinor the Green's  function (GF) matrix in Matsubara representation,
  is obtained via
$G_{ {\bf  k}{\bf  k}^\prime}(\tau)=-\langle T \hat{\psi}_{{\bf k} }(\tau)
\hat{\psi}_{{\bf k}^\prime}^{\dagger}(0)\rangle$,
whence
\bea
% \lefteqn{
G_{ {\bf  k}{\bf  k}^\prime}(\omega_n)
%} && \nonumber\\&&
=G^{0}_{{\bf k}}(\omega_n)[\delta_{ {\bf  k}{\bf  k}^\prime} +
{\hat T}_{ {\bf  k}{\bf  k}^\prime}(\omega_n)G^{0}_{{\bf k}^\prime}(\omega_n)],
% \nonumber\\
\label{GF}
\eea
where $G^{0}_{{\bf k}}(\omega_n)=\left( i\omega_n-\hat{\beta}_{{\bf k}} \right)^{-1}$ is
the unperturbed propagator (bare GF) of the conduction electrons. Its poles are given by the cf-hybridized (due to $V_{{\bf k}}$) quasiparticle excitations $E^{\pm} _{\bf  k}$ gapped by $\Delta_{{\bf k}}$ which are contained in the $\hat{\beta}_{{\bf k}}$ matrix.
As a result of the hybridisation gap of order $\tilde{V}_{\bf k}$ the QPI interference pattern will depend strongly on the energy or bias voltage within the gap range.

Solving the Dyson equation for the ${\hat T}$-matrix
\be {\hat T}_{ {\bf  k}{\bf  k}^\prime}(\omega_n)=\hat{U}_{ {\bf  k}{\bf  k}^\prime}+\sum_{{\bf k}^{\prime \prime}} \hat{U}_{ {\bf  k}{\bf  k}^{\prime \prime}}G^{0}_{{\bf k}^{\prime\prime}}(\omega_n)
 {\hat T}_{{\bf k}^{\prime \prime}{\bf k}^{\prime}}( \omega_n)
,\ee
the LDOS can be obtained from imaginary part of the full GF,
via analytic continuation
$i\omega_n\rightarrow \omega+i  0^{+}$ according to
\bea
N^{c}_{{\bf r}}(\omega)&=&\frac{-1}{\pi}
\mbox{Im Tr} \,\left[  (\frac{\tau_0+\tau_z}{2})G(r,r,\omega_n)\right]_{i\omega_n\rightarrow \omega+i 0^+}
\nonumber\\
&=& N^{0}_{{\bf r}}(\omega)+\delta N^c_{ {\bf  r}}(\omega).
\label{DOS}
\eea
The  Fourier transform of
the fluctuations, $\delta N^c_{ {\bf  k}}(\omega)$, can be obtain using Eq.~(\ref{GF})
in following form,
\bea
&&\delta N^c_{ {\bf  k}}(\omega)=-\frac{1}{\pi}\mbox{Im
Tr}
\\\nonumber
&&
\sum_{{\bf k}}
 \left[  (\frac{\tau_0+\tau_z}{2})
G^{0}_{{\bf k}}(\omega_n)
{\hat T}_{ {\bf  k}{\bf  k}+{\bf  q}}(\omega_n)G^{0}_{{\bf k}+{\bf  q}}(\omega_n)
\right]_{i\omega_n\rightarrow \omega+i 0^+}
\eea
This quantity is called QPI and
qualitatively
 is proportional to the convolution of density of states (DOS) contributions at the initial and final states momenta of
the Brillouin zone which lie on a surface  of constant  energy\cite{Balatsky06} , i.e.,
\be
\delta N^c_{ {\bf  q}}(\omega)\propto
\int N^{0}_{ {\bf  k}}(\omega)N^{0}_{ {\bf  k} +{\bf  q}}(\omega)d {\bf  k},
\ee
where $ N^{0}_{ {\bf  k}}(\omega)$ is the spectral density function (DOS contribution) originating at momentum \bk~ at given bias   energy $\omega$.
%%%%%%%%%%%%%%%%%%%%%% figure %%%%%%%%%%%%%%%%%%%%
\begin{figure}
\centerline{
\includegraphics[width=0.99\linewidth]{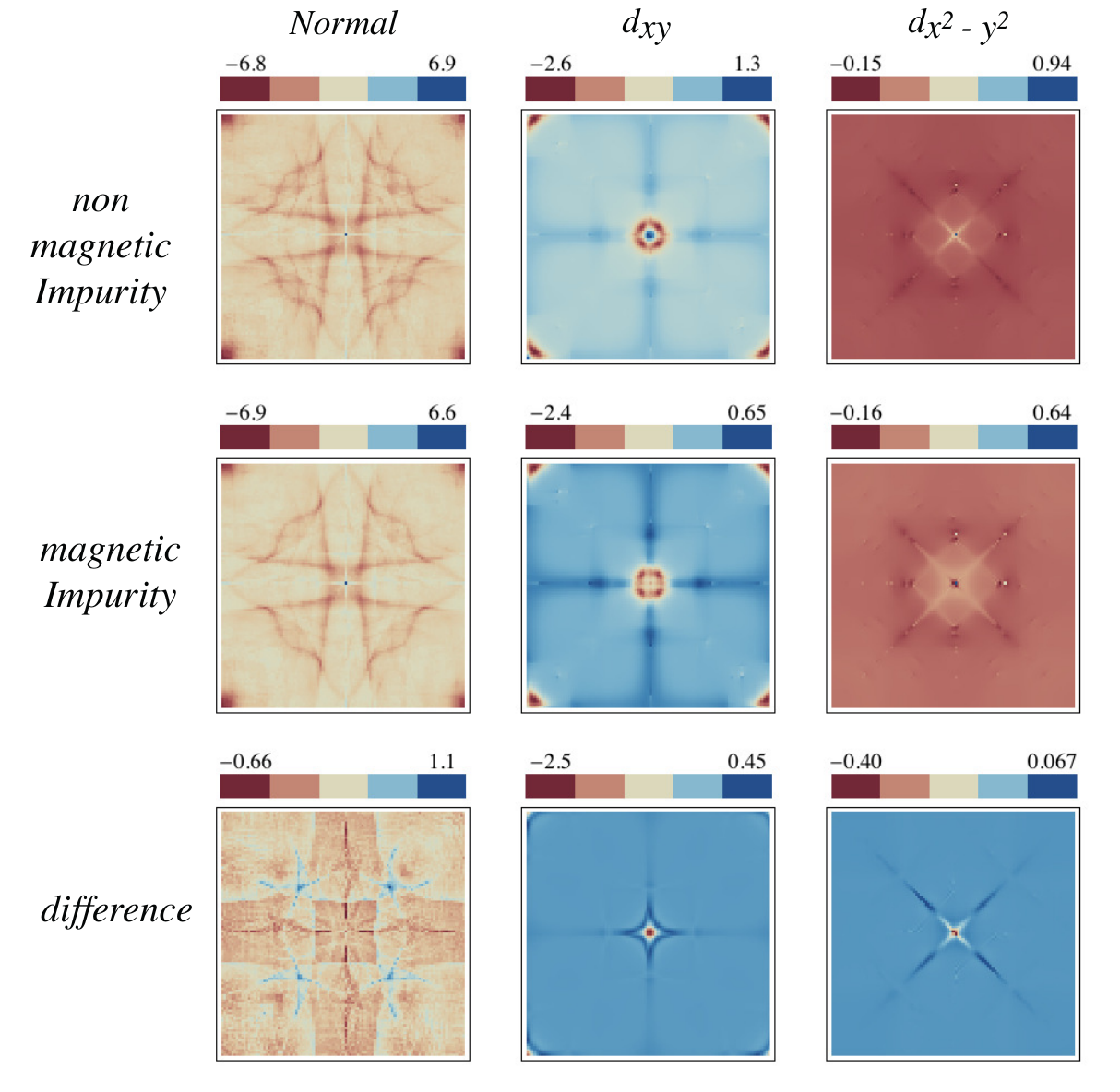}
}
\caption{(Color online) Total quasiparticle interference  (quasiparticle interference in Fig.~\ref{Difkz} averaged over $k_z$),
for normal state (first panel) , superconducting state with $d_{xy}$ gap  symmetry (second panel)
 and superconducting state with  $d_{x^2-y^2}$ gap  symmetry (third panel). (For bias voltage $\omega=0.01\Delta_0$ and $\Delta_0/W=0.086$).
First row corresponds to the non-magnetic impurity,
second row refers to magnetic impurity and last one shows  the difference $\delta QPI$
between quasiparticle interference for magnetic  and non-magnetic impurity scattering.}
\label{Totkz}
\end{figure}
%%%%%%%%%%%%%%%%%%%%%%fig%%%%%%%%%%%%%%%%%%%%%%%%%

%%%%%%%%%%%%%%%%%%%%%% figure %%%%%%%%%%%%%%%%%%%%
\begin{figure}
\centerline{
\includegraphics[width=0.99\linewidth]{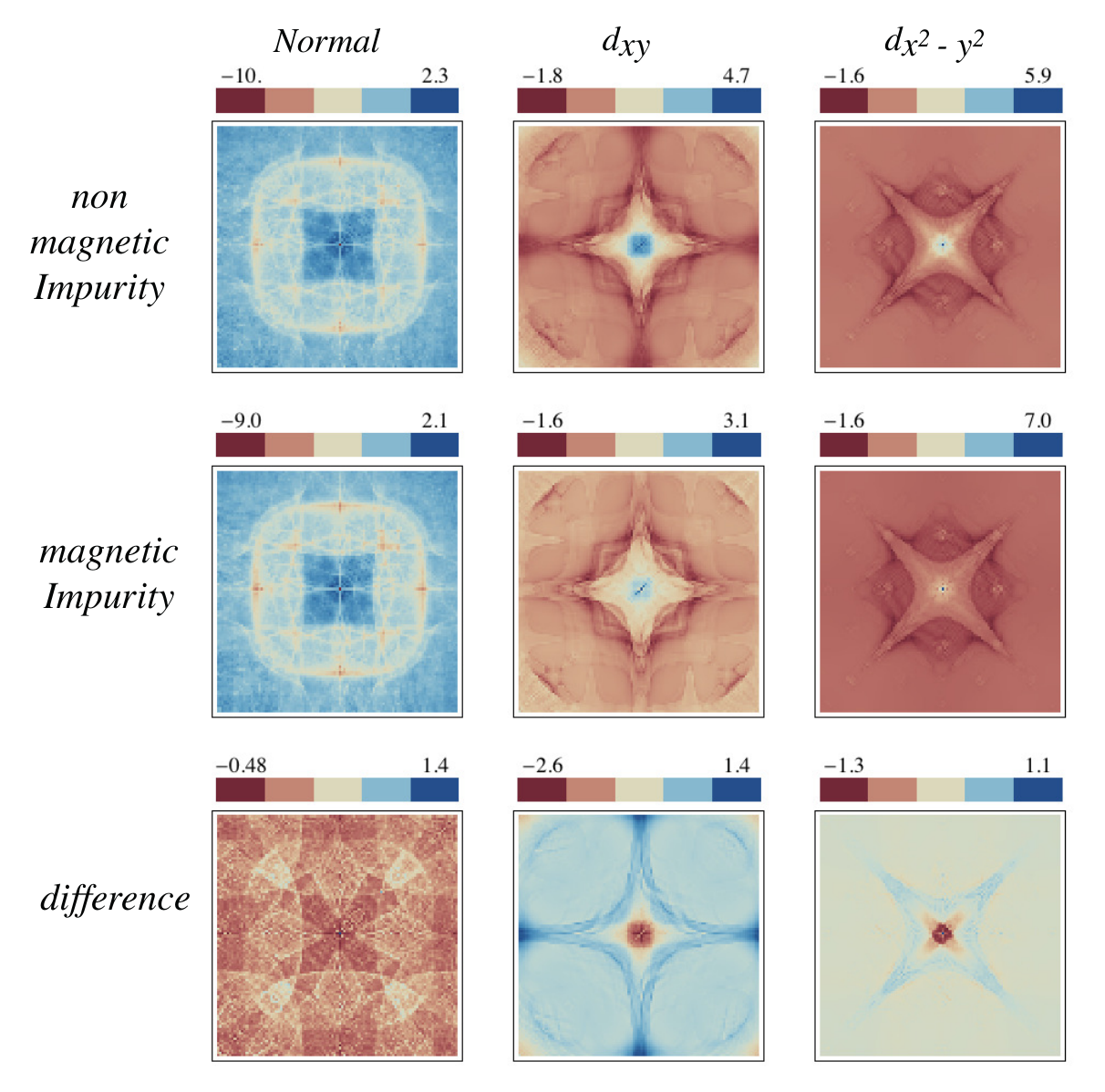}
}
\caption{(Color online) Same as Fig.~\ref{Totkz} but for bias voltage $\omega=0.1\Delta_0$.}
\label{TotkzL}
\end{figure}
%%%%%%%%%%%%%%%%%%%%%%fig%%%%%%%%%%%%%%%%%%%%%%%%%

%%%%%%%%%%%%%%%%%%%%%%%%%%%%%%%%%%%%%%%%%%%%%%%%%%%%%%%%%%%%%%%%%%%%%%%%%%
%%%%%%%%%%%%%%%%%%%%%%%%%%     Section IV     %%%%%%%%%%%%%%%%%%%%%%%%%%%%
%%%%%%%%%%%%%%%%%%%%%%%%%%%%%%%%%%%%%%%%%%%%%%%%%%%%%%%%%%%%%%%%%%%%%%%%%%
\section{Numerical results}
\label{numerical}
Now  we start to examine the effect of the single impurity scattering center on the quasiparticle interference pattern of  $\rm{ CeCoIn_5}$.
Since the corrugated Fermi surface of $\rm{ CeCoIn_5}$ has considerable three dimensional character,
we are not able to use the ${\hat T}$-matrix formalism directly to compare with  the  two dimensional (surface)  FT-STM results.
In this respect we  restrict ourself to the ab plane by averaging over the momenta in  the $k_z$ direction.
Using the Eqs.~(\ref{GF}-\ref{DOS}) one can calculate the  ${\hat T}$-matrix and corresponding Green's function for each slice of the Fermi surface at a particular $k_z$,  then by averaging over the individual QPI of each  $k_z$ slice it is easy to  find the
total QPI for tunneling current along (001)-direction.\\

The result of  the QPI for the non-magnetic impurity is presented in Fig.~\ref{Difkz}, for individual layers at $k_z c=0,0.2\pi,0.4 \pi, 0.6 \pi, 0.8 \pi, \pi$.
The first panel of this plot shows  the  corresponding quasiparticle interference strength in the  normal state, and
to compare the different superconducting gap symmetries in this compound, we
look at the two main candidates discussed above, namely \\
$d_{xy}$ gap  symmetry,
$$
\Delta_{{\bf k}}=\Delta_0 \sin k_x a \sin k_y a,
$$
and  $d_{x^2-y^2}$ gap  symmetry,
$$
\Delta_{{\bf k}}=\Delta_0 ( \cos k_x a- \cos k_y a),
$$
 in the second and third panels respectively.
  In Fig.~\ref{Fermi}(b) the constant energy intensity maps of the spectral density for normal state (first panel)
    superconducting state with $d_{xy}$ (second panel) and  $d_{x^2-y^2}$(third panel) gap  symmetry
    are plotted for the different cuts of the Fermi surface at $k_z c=0, 0.5
    \pi, \pi$.
Their node structure with respect to the Fermi surface is indicated by dashed lines.

First we analyze the results for individual $k_z$ cuts for nonmagnetic impurities.
One can see  that for the  cases with $k_z c=0,0.2\pi, 0.4 \pi$,  the gap with  $d_{xy}$ symmetry does not have  a node on the Fermi surface.
Since the bias voltage is smaller than the gap value, DOS is reduced and as a result the QPI becomes negligible.
For larger $k_z$ (because of the inner structure of the spectral density function) when the $d_{xy}$ gap becomes nodal or the node-lines approach the Fermi surface
a different behavior of QPI appears. An enhancement of QPI for the distinguished \bq-vector which joins the new node points is observed
for  $k_z c=0.6\pi,0.8\pi$ (see $\bq_1$ and $\bq_2$  in Fig.~\ref{Fermi}(b.V)),
by approaching to  $k_z c=\pi$ the size of the inner part is growing thus the effect of the node
and QPI become smaller again.

This scenario  is completely different for the $d_{x^2-y^2}$ case where the node lines cut the Fermi surface at any value of $k_z$.
Because in the regions close to the node line, the DOS contributions of the scattered electrons
 are rapidly growing, QPI will be enhanced for the connecting  $\bq_i$ - vectors, and a corresponding point like pattern at  $\bq_i$
  can be found in all values of $k_z$.
Especially when the inner part of spectral density function appears the number of these points, as shown in last panel of the Fig.\ref{Fermi}b, are increased.

The observable tunneling current is an average over the Fermi surface slices
at different $k_z$. Therefore we have to consider the calculated QPI averaged in (001)-direction.
For this reason we plot in Fig.~\ref{Totkz} the
averaged QPI pattern,
for normal state in the first panel, and superconducting state with $d_{xy}$
 and  $d_{x^2-y^2}$ gap  symmetry in second panel and third panel, respectively.

In this figure the first row corresponds to QPI from non-magnetic impurity scattering
which is calculated by averaging over the individual $k_z$- layers presented in Fig.~(\ref{Difkz}).
In the normal state a trace of the dominant $\bq$ - vectors which join the prominent points of the spectral density functions (Fig.~\ref{Fermi}.b  first panel) can still be followed. The QPI is most pronounced at  small $\bq$  - vectors and also around  the zone boundary vectors ($\pm\pi, \pm\pi$).

In the superconducting state with $d_{xy}$ gap symmetry
due to the gap opening  the strength of the QPI decreases, and if we neglect the inner structure of spectral density  the remaining part has very small effect in QPI.
By considering the inner part of the spectral density function
which has a node, pronounced QPI  pattern can be found at corners   ($\pm\pi, \pm\pi$) and also in (100)- and (010)-directions.
For a $d_{x^2-y^2}$ gap symmetry  QPI pattern in the Brillouin zone is distinctly different from normal state as well as $d_{xy}$ state.
In this regime the
pronounced points QPI  pattern can be found at small $\bq$ vectors in (110)- , (010)-and (100)-directions.

The second row of Fig.~\ref{Totkz}  refers  to QPI strength due to magnetic impurity scattering.
The general behavior for the magnetic impurity is the same as the non magnetic impurity.
Consequently we do not see dramatic change of QPI pattern aside from an overall reduction of the QPI amplitude at small momenta.
Instead we focus on the most interesting QPI quantity for a comparison with possible experimental results which is the {\it difference} of QPI strength for the magnetic (m) and non-magnetic (n.m) impurity scattering defined by $\delta {\rm QPI}={\rm QPI_m}-{\rm QPI_{n.m}}$. We plot this difference in the last row of the Fig.~\ref{Totkz}.
Magnetic impurity scattering leads to enhanced QPI
 for the large  \bq - vectors,  and  to  reduced QPI
for the small \bq - vectors, in the superconducting states.
The QPI in the superconductor with $d_{xy}$ gap symmetry is along the (100)-and (010)-directions and
there are some
effects at ($\pm\pi, \pm\pi$),  but for  the $d_{x^2-y^2}$ gap symmetry we observe  small $\bq$ vector structures along the diagonal
and we do not see any effect at large vectors around  ($\pm\pi, \pm\pi$).

By increasing the bias voltage size the small pockets in spectral density function are growing which  enhances  these effects.
In the Fig.~\ref{TotkzL} we present the QPI for the larger bias voltage  $\omega=0.1\Delta_0$.
As we expect by increasing the size of the electronic pockets, the QPI pattern grows, and the intensity  shifts to higher momenta.
 The QPI  still has an axis-aligned cross structure for $d_{xy}$ and diagonal cross  continuous pattern with sharp points along (100)- and (010)-directions for $d_{x^2-y^2}$.
By increasing the bias voltage the QPI patterns which are  created by the inner part of spectral density function  are growing for both symmetries.
However
 the sharp
 point like picture along (100)- and (010)-directions in for the $d_{x^2-y^2}$  and  at ($\pm\pi, \pm\pi$) for  $d_{xy}$ are  still
 distinguishable.

Finally we remark on the effect of a magnetic field acting only on the spin degrees of freedom.
Our calculations show that  adding the Zeeman term ${\cal H}_B=\sigma\mu_B B$  to the quasiparticle dispersion
does not effect the previous result dramatically for accessible field strength.
Another more promising direction would
be to study the influence of a FFLO type spatially inhomogeneous superconducting order parameter on QPI pattern in the high field regime where the FFLO state has been identified e.g. in recent NMR experiments \cite{kumagai:11}.

%%%%%%%%%%%%%%%%%%%%%%%%%%%%%%%%%%%%%%%%%%%%%%%%%%%%%%%%%%%%%%%%%%%%%%%%%%
%%%%%%%%%%%%%%%%%%%%%%%%%%     Section IV     %%%%%%%%%%%%%%%%%%%%%%%%%%%%
%%%%%%%%%%%%%%%%%%%%%%%%%%%%%%%%%%%%%%%%%%%%%%%%%%%%%%%%%%%%%%%%%%%%%%%%%%
\section{Summary}
\label{summary}

In this paper we presented the theory for the SI-STM  in the heavy Fermion superconductor.
We study the  quasiparticle interference  in heavy Fermion superconductor $\rm{ CeCoIn_5}$ systematically.
We compare the effect of the  different  singlet superconducting  order parameters that have been proposed, namely $d_{xy}$ and
$d_{x^2-y^2}$ , on the  QPI in  this compound.
By comparing the QPI pattern of  single non-magnetic and magnetic impurities
we have shown that
QPI has an axis aligned cross structure with some features at  ($\pm\pi, \pm\pi$)  for $d_{xy}$ gap function,
and a diagonal cross continuos pattern with sharp points along (100)- and (010)-directions for $d_{x^2-y^2}$ gap symmetry.
Furthermore we have shown the difference of QPI in present of magnetic and non-magnetic impurity, $\delta$QPI, becomes positive for large and negative for small wave vectors. We have shown that the  axis aligned and diagonal cross  structures are clearly  seen in the $\delta$QPI for $d_{x^2-y^2}$  and   $d_{xy}$ gap symmetries respectively.

We conclude that QPI pattern are sensitive to the order parameter symmetry in the 115 compounds and may be observed in future STM measurements. This observation would provide a direct fingerprint to distinguish between different superconducting order parameter models. It may be particularly relevant for the  case of  $\rm{ CeIrIn_5}$ where the symmetry of the superconducting gap function is discussed controversially \cite{shakeripour:07,kasahara:08}.

%%%%%%%%%%%%%%%%%%%%%%%%%%%%%%%%%%%%%%%%%%%%%%%%%%%%%%%%%%%%%%%%%%%%%%%%%%
%%%%%%%%%%%%%%%%%%%%%%%%%%%%%      References        %%%%%%%%%%%%%%%%%%%%%
%%%%%%%%%%%%%%%%%%%%%%%%%%%%%%%%%%%%%%%%%%%%%%%%%%%%%%%%%%%%%%%%%%%%%%%%%%

\end{document}